\documentclass[
aps,
prd,
amsmath,
amssymb,
amsfonts,
floatfix,
nofootinbib,
superscriptaddress,
longbibliography,
10pt,
bibnotes,
reprint,
noeprint]{revtex4-2}

\usepackage[english]{babel}
\usepackage[utf8]{inputenc}
\usepackage[T1]{fontenc}

\usepackage[dvipsnames]{xcolor}
\usepackage{tikz}
\usepackage{CJKutf8}
\usepackage{verbatim}
\usepackage{graphicx}
\usepackage[all]{xy}
\usepackage{simplewick}
\usepackage{setspace}
\usepackage{array}
\usepackage{enumitem}
\usepackage{bbm}
\usepackage{bm}
\usepackage{mathtools}
\usepackage{subfigure}
\usepackage{physics}
\usepackage{mathrsfs}
\usepackage{enumerate}

\usepackage[
colorlinks,
linkcolor=BrickRed,
citecolor=MidnightBlue,
urlcolor=MidnightBlue
]{hyperref}

\allowdisplaybreaks[4]

\newcommand{\wtk}{\widetilde{k}}

\renewcommand{\tilde}{\widetilde}

\newcommand{\edoc}{\end{document}}
\newcommand{\eref}[1]{(\ref{#1})}

\newcommand{\nn}{\nonumber}

\renewcommand{\i}{\mathrm{i}}

\renewcommand{\mod}{\mathrm{\,mod\,}}

\newcommand{\be}{\begin{eqnarray}}
	\newcommand{\bea}{\begin{eqnarray}}
		\newcommand{\eea}{\end{eqnarray}}
	\newcommand{\beq}{\begin{equation}}
		\newcommand{\ee}{\end{eqnarray}}
	\newcommand{\eeq}{\end{equation}}

        \newcommand{\bmat}{\left ( \begin{array}{cc} }
	\newcommand{\emat}{\end{array} \right ) }
	\newcommand{\wk}{\widetilde k}
	\newcommand{\wl}{\widetilde l}
	\newcommand{\wm}{\widetilde m}
	\newcommand{\wn}{\widetilde n}

\newcounter{amg}

\newcounter{ls}

\newcounter{jvc}

\newcounter{jz}

\begin{document}

\title{Many-Body Quantum Chaos At All Time Scales}

\author{Antonio  M. Garc\'\i a-Garc\'\i a}
\email{amgg@sjtu.edu.cn}
\affiliation{Shanghai Center for Complex Physics,
	School of Physics and Astronomy, Shanghai Jiao Tong
	University, Shanghai 200240, China}

\author{Lucas S\'a}
\email{ld710@cam.ac.uk}
\affiliation{TCM Group, Cavendish Laboratory, University of Cambridge, JJ Thomson Avenue, Cambridge CB3 0US UK}

\author{Jacobus J. M. Verbaarschot}
\email{jacobus.verbaarschot@stonybrook.edu}
\affiliation{Center for Nuclear Theory and Department of Physics and Astronomy, Stony
  Brook University, Stony Brook, New York 11794, USA}

\author{\begin{CJK*}{UTF8}{gbsn}%
		Jie-Ping Zheng (郑杰平)
\end{CJK*}}
\email{jpzheng@sjtu.edu.cn}
\affiliation{Shanghai Center for Complex Physics,
	School of Physics and Astronomy, Shanghai Jiao Tong
	University, Shanghai 200240, China}

\begin{abstract}
  We describe the dynamics of many-body quantum chaotic systems at all time scales by studying the Green's and out-of-time-order correlation (OTOC) functions of the four-body, $N$-Majorana Sachdev--Ye--Kitaev model. By combining the scramblon formalism and random-matrix-theory techniques, we obtain analytical expressions for these functions at all times. The early exponential growth of the OTOC is followed by 
  an exponential decay at a rate governed by that of the Green's function
  (the real part of the leading complex Ruelle--Pollicott resonances).
 For late times that scale exponentially with $N$, both functions have a dip-ramp-plateau pattern for  $N\mod8 = 2, 6$ that deviates substantially from the ergodic prediction due to local-in-energy correlations of matrix elements and eigenvalues, even after the Heisenberg time.
  \end{abstract}

\maketitle
Quantum chaos has played an important role in a broad variety of research fields, from the description of correlations of  excitations in nuclei \cite{wigner1951,aberg1990} and the Euclidean
dynamics of quarks in lattice QCD \cite{verbaarschot1993a,shuryak1993a} to the real time dynamics of
electrons and waves in random media \cite{bohigas1984,larkin1969,efetov1983},
statistical physics \cite{leboeuf2002}, and superconductivity \cite{olofsson2008,garcia2008}. It has also sparked
research in semiclassical techniques \cite{gutzwiller1971} and
random matrix theory (RMT) \cite{mehta2004,cipolloni2024,cipolloni2023,guhr1998}.
More recently, the relations between quantum chaos and low-dimensional
quantum gravity \cite{sekino2008,shenker2014,kitaev2015,maldacena2016,maldacena2015} have stimulated both fields. 

This broad interest in quantum chaos is motivated by the universality of its predictions, such as the exponential growth
of quantum uncertainty for short times and RMT spectral correlations, signaling quantum ergodicity, for long times.
Based on gravitational ideas, a universal bound on the exponential growth of quantum uncertainty was proposed \cite{maldacena2015},
which is saturated in theories with a gravity dual. Kitaev found such an example \cite{kitaev2015}, the so-called Sachdev--Ye--Kitaev (SYK) model \cite{bohigas1971,french1970,sachdev1993,benet2003,kitaev2015,maldacena2016}
consisting of $N$ Majoranas in zero dimensions with a $q$-body random interaction of infinite range in Fock space.
Therefore, it was proposed that the SYK model has a gravity dual, which, shortly thereafter, was identified \cite{maldacena2016a} as Jackiw--Teitelboim (JT) gravity \cite{jackiw1985,teitelboim1983}. Using RMT techniques, it was shown
\cite{garcia2016,cotler2016} that the SYK model is quantum chaotic at time scales beyond $\log N$,
while JT gravity is quantum chaotic because its field theory dual in this limit is a random matrix \cite{saad2019,garciagarcia2019,Altland:2024ubs}. 

In this Letter, we study many-body quantum chaos at all time scales, which is characterized by two well-separated large parameters, the number of particles, $N$, and the dimension of the Hilbert space, $\exp N$. 
In many-body systems with no classical limit, such as the SYK model, chaos cannot be defined by correspondence with classically chaotic dynamics; we rely instead on
intrinsically quantum diagnostics, such as RMT spectral correlations and
the growth of operator complexity \cite{Roberts:2018mnp} under Heisenberg evolution (measured by the length of a string of Majorana operators).
For $q>2$, the length of an operator increases exponentially with time until it saturates 
to the ergodic limit where all operators are equally likely. This happens at a time scale of order $ \log N$. Indeed, for the connected spectral form factor (SFF), the onset of RMT
behavior may be as early as $t\sim \log N$ \cite{Gharibyan:2018jrp,Altland:2017eao,Jia:2019orl}.
Operator complexity can also be measured by the average square
of the anticommutator of elementary operators \cite{Roberts:2018mnp}, a quantity closely related to out-of-time-order correlation (OTOC) functions. For simplicity, we will refer to them generically as OTOCs \cite{larkin1969,berman1978,jalabert2018,garciamata2023,vallejo2025}.
Exponential growth of the OTOC alone does not establish chaos, as it also occurs near unstable fixed points of systems with a regular classical limit~\cite{hummel2019,pilatowsky2020,scaffidi2019}. Such false positives are
exposed by the long-time behavior of the OTOC, which differs qualitatively from the chaotic case~\cite{fortes2019,kidd2021,garciamata2023}; this motivates us to study the OTOC at all time scales.

For an integrable ($q=2$) SYK model, the OTOC---derived analytically for all times in Ref.~\cite{garciagarcia2025}---does not experience exponential growth and decays as a power law related to the Fourier transform of the single-particle spectral density. RMT behavior is found for $t >\sqrt N$.
Other  results for OTOCs were obtained by using various methods and limits, exploiting duality \cite{shenker2014,maldacena2016a,shenker2015,giombi2023,choi2023} to JT gravity using the Dray-'t Hooft shockwave formalism \cite{dray1985,verlinde1992}, the Schwarzian limit of the SYK model \cite{stanford2017,bagrets2017, yang2019,kitaev2019,mertens2017}, the large-$q$ limit \cite{kitaev2018,streicher2020,gu2019a,gu2022}, the Brownian SYK \cite{stanford2022,cirac2019}, RMT \cite{cotler2017,cipolloni2024,torresherrera2017},
and the double-scaled SYK (DSSYK) model  \cite{berkooz2018,berkooz2019,berkooz2024} (see Ref.~\cite{mertens2023} for a recent review).
Significant progress on OTOCs for general $q$ was made in Ref.~\cite{gu2022}, where the calculation was reduced to numerically solving the Schwinger--Dyson (SD) equations for an out-of-time-order source.
 
\begin{figure}[t]
	\begin{tabular}{@{}r@{}}
    \includegraphics[width=0.74\columnwidth]{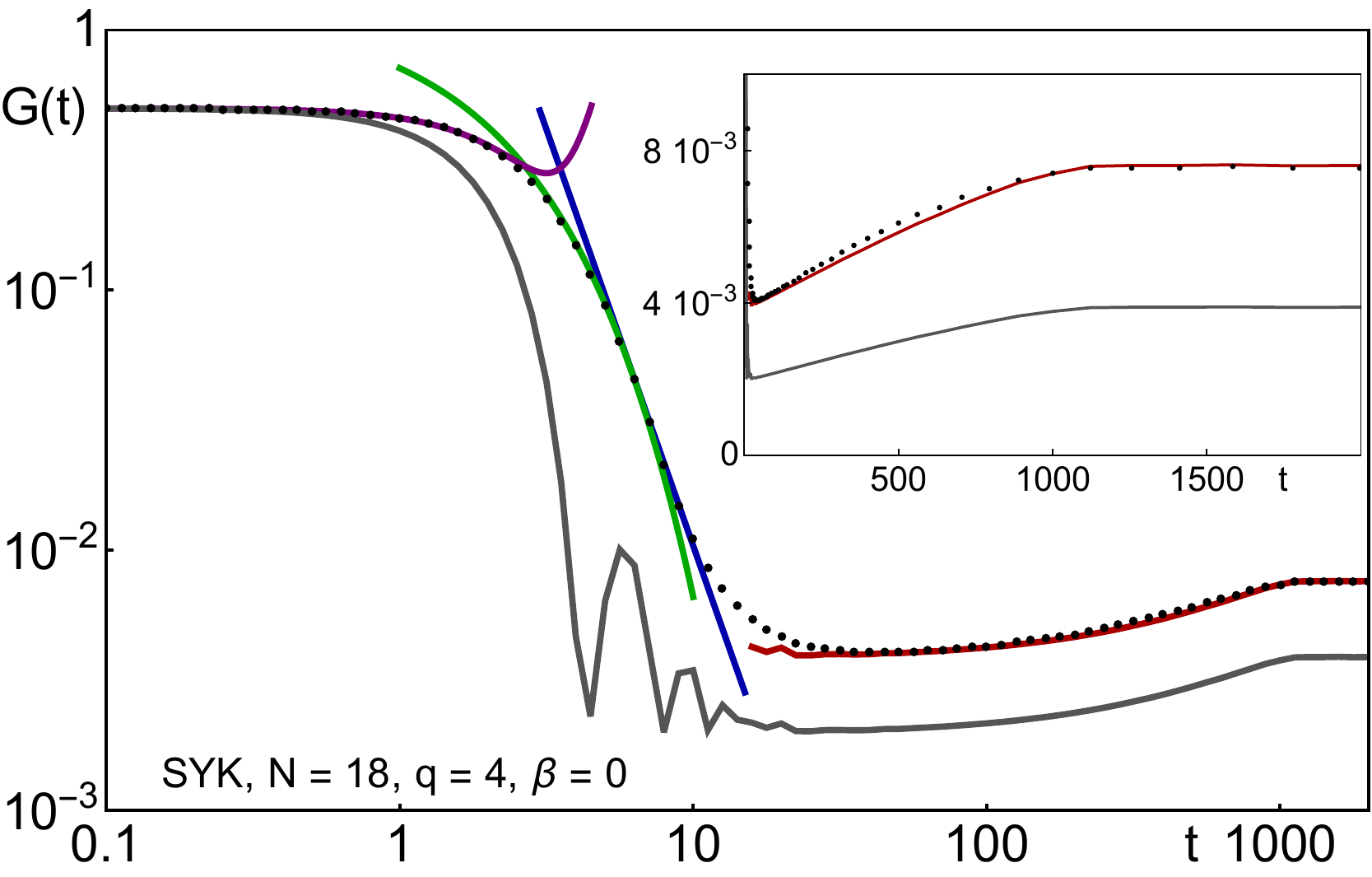}\\
	\includegraphics[width=0.755\columnwidth]{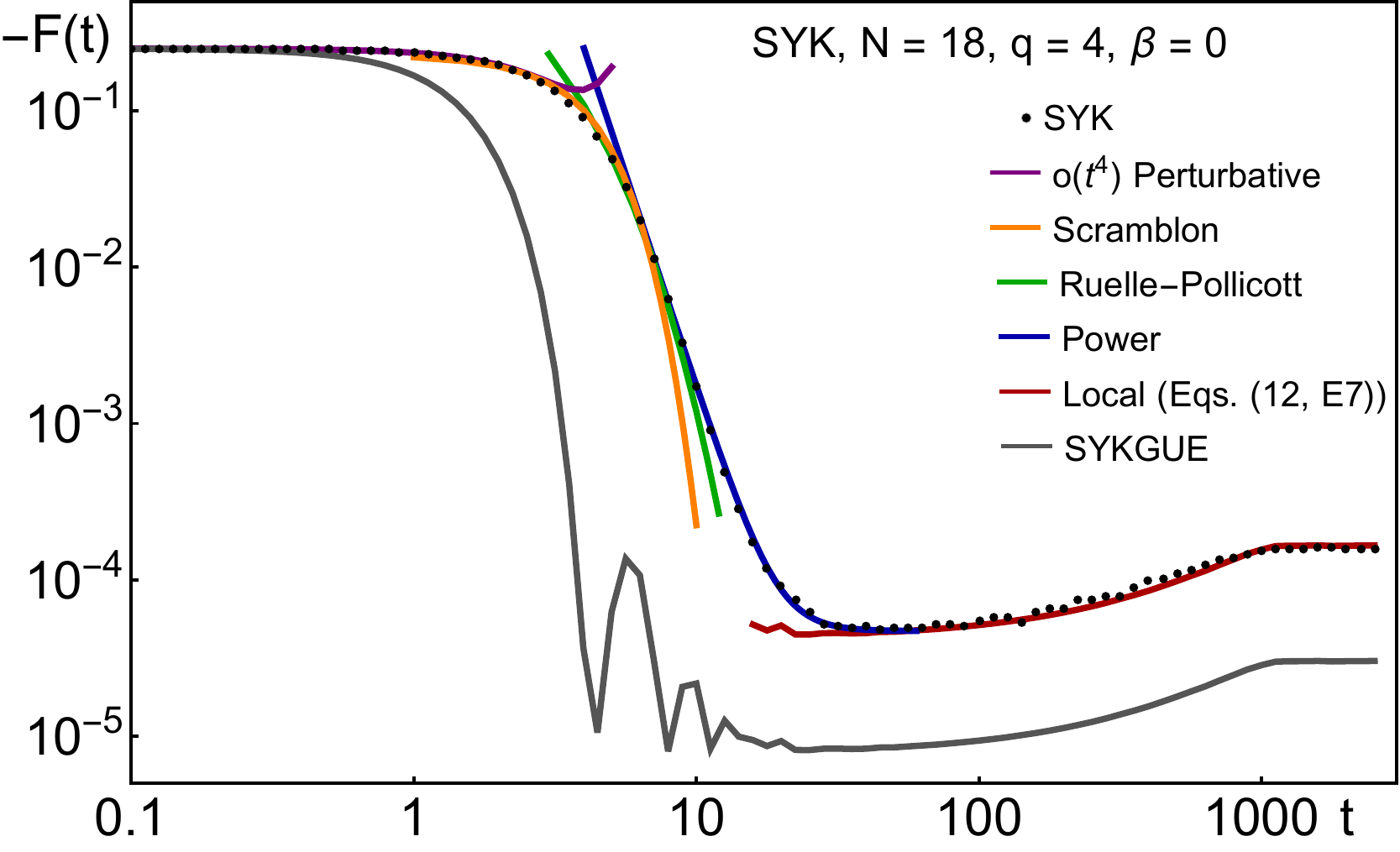}
	\end{tabular}
	\caption{Time dependence of the Green's function (top) and  OTOC (bottom) on a log-log scale. Numerical results (black dots) obtained from exact diagonalization for the $N=18,\; q=4, \; \beta=0$ SYK model are compared to  numerical fits and analytical results derived below, given by solid curves as indicated in the legend of $F(t)$, also for $G(t)$ (see Appendix \ref{app:fit} for the fitting parameters). The inset in the top panel compares the local prediction and the SYK results on a linear scale and better shows the discrepancy of less than 4\% for $t>50$. See Appendix~\ref{app:N26} for $N=26$.}
    \label{fig:all}
\end{figure}

Although much progress has occurred in the past few years, we are still far from a comprehensive understanding of all time scales that govern the dynamics of many-body quantum chaotic systems. The main goal of this Letter is to fill this gap starting from the perturbative scale and extending beyond the Heisenberg time.
We focus exclusively on the $q = 4$ SYK model, which is more challenging than the Brownian or large-$q$ limits, but also more realistic because two-body interactions are ubiquitous in nature. Indeed, the  $q=4$ (complex) SYK model was first introduced as a model for the nuclear
two-body interaction \cite{French1971,bohigas1971}.
The presented results are for the infinite-temperature limit---the temperature dependence of the results is mostly quantitative. 

{\it Domains and time scales of many-body quantum chaos.---}%
We identify five different scaling domains for the OTOC: the perturbative domain, the exponential-growth (scrambling) domain, the exponentially decaying (relaxation) domain characterized by Ruelle--Pollicott (RP) resonances \cite{ruelle1986,pollicot1985,prosen2002,maldacena2016a,shenker2014,polchinski2015,jalabert2018,garciamata2023,mori2024,yoshimura2025theory,vallejo2025}, the power-law domain, and the ergodic domain characterized by RMT. Using the scramblon approach of Ref.~\cite{gu2022}, which had not previously been applied to the OTOC of the $q=4$ SYK model, we obtain
analytical results for the exponential-growth and exponential-decay domains; this is the first major result of this Letter.
The second major result is the parameter-free expression for the  RMT domain based on the insight that the phase $\exp\{it(E_k -E_l)\}$ limits
significant contributions to the OTOC and the Green's function to terms with SYK energies $E_k\approx E_l$. This result is strongly enhanced with respect to the RMT expectation.
A related important finding is that power-law behavior due to the edge of the spectrum is also strongly enhanced with respect to RMT results.
Our analytical expressions have been confirmed by finite-$N$ exact diagonalization and Krylov subspace techniques for graphics processing units (GPUs) \cite{garcia2023d} that enable us to reach $N \leq 50$.

{\it  Models and correlation functions.---}%
The $q=4$ SYK Hamiltonian \cite{kitaev2015} for $N$ interacting Majorana fermions $\gamma_i$, termed SYK from now on, is given by
\begin{equation}
	H = \sum_{i<j<k<l}^N J_{ijkl} \gamma_i \gamma_j \gamma_k \gamma_l,
\end{equation}
where $\{\gamma_i, \gamma_j\} = \delta_{ij}$ and each $J_{ijkl}$ is sampled from a Gaussian distribution with zero mean and variance $6N^{-3}$.
The Hilbert space dimension is $D=2^{N/2}$. 

The Green's function $G(t)$ is defined as
\begin{equation}
\label{eq:green}
  G(t)=
  \frac 1D \tr \left[e^{(it-\beta/2)H}\gamma_{i}e^{-(it+\beta/2)H}\gamma_{i}\right],
\end{equation}
where $\beta$ is the inverse temperature (which we set to $\beta=0$) and $i = 0, \dots,  N-1$. 
It is independent of $i$ and can thus also be replaced by
an average over $i$.
The OTOC is defined as ($i \ne j$)
\begin{equation}
	\label{eq:otoc}
 F(t)=\frac 1D \tr \left[(e^{(it-\beta/4)H}\gamma_{i}e^{-(it+\beta/4) H}\gamma_{j})^2\right].
\end{equation}
Instead of keeping $i$ and $j$ fixed, we can again average over these indices. 

\begin{figure}[t]	
  \begin{tabular}{@{}l@{}}
  \includegraphics[width=0.74\columnwidth]{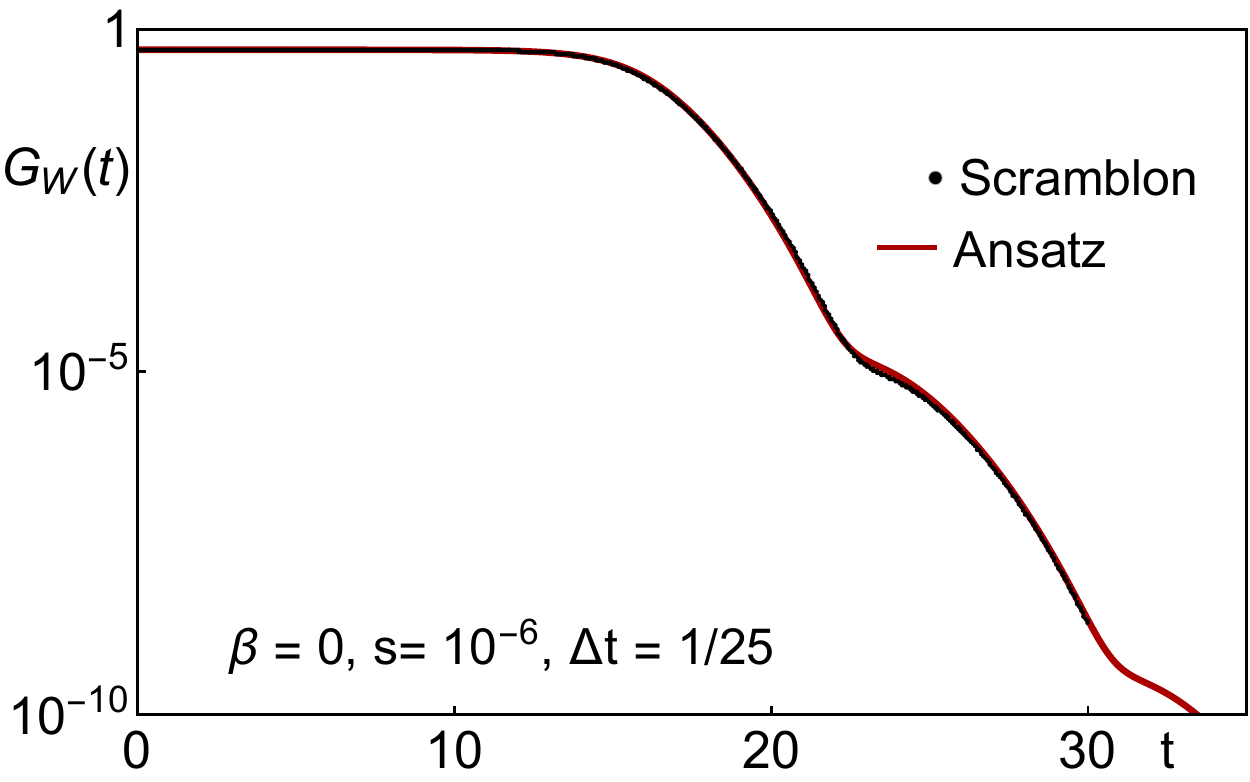}\\
  \includegraphics[width=0.755\columnwidth]{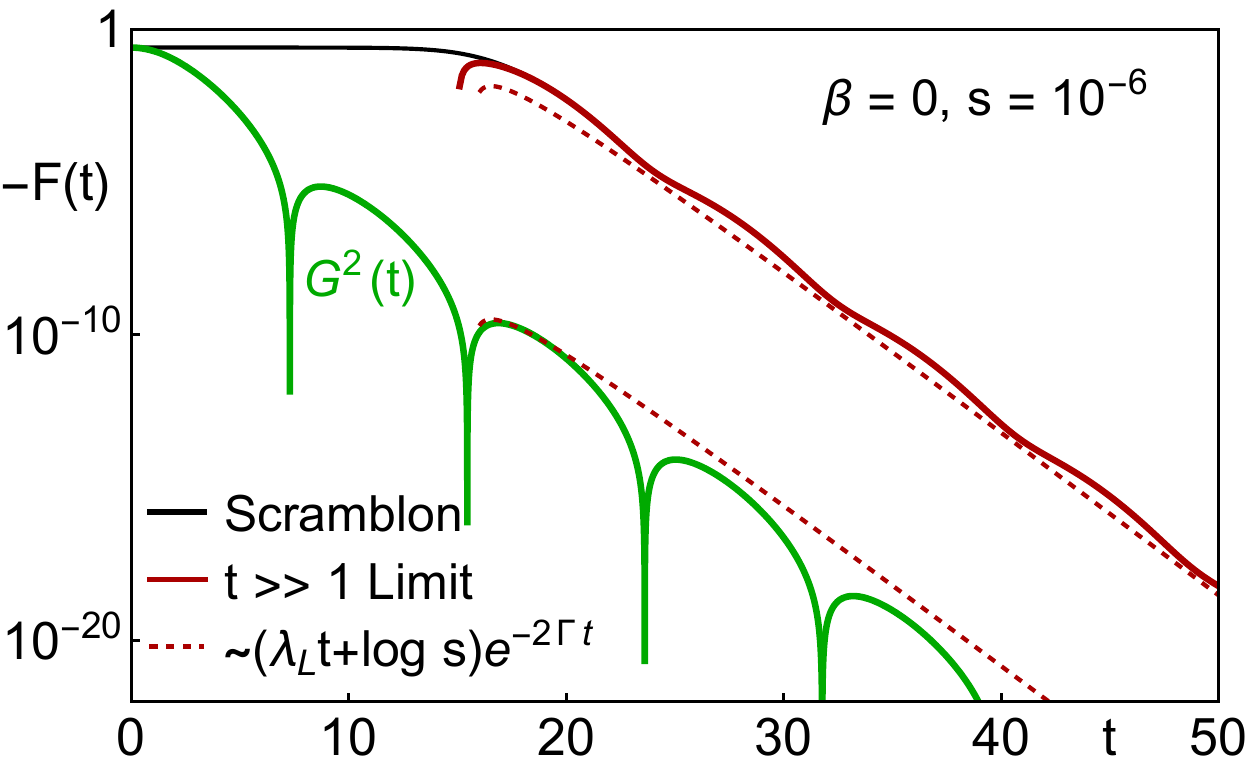}
  \end{tabular}
  \caption{
		Top panel: The Wightman function $G_W(t)$ at $\beta=0$ for an out-of-time-order source with strength $s=10^{-6}$, obtained from  the SD equations on the Keldysh contour with spacing $\Delta t=0.04$, following the methods outlined in Appendix~C of Ref.~\cite{gu2022} (black), compared to the ansatz Eq.~\eref{gwan} with one fitting parameter (red). 
        For large $t$, $G_W(t) \sim a \exp[-2\Gamma t] + {\rm Re}
        ( b \exp[-2(\Gamma+i\Omega )t])$ shows weaker oscillations than
        the squared Green's function $G^2(t) \sim  {\rm Re}( \tilde b \exp[-2(\Gamma+i\Omega )t])$ obtained by solving the SD  equations.
        Bottom panel: The OTOC, $-F(t)$, obtained from the scramblon method combined with the ansatz Eq.~\eref{gwan} (black), compared to  $G^2(t)$ (green) and the large-$t$ limit (red), $F(t) \sim (\lambda_L t +\log s) \exp(-2\Gamma t)$ (dashed) and $F(t) \sim (\lambda_L t +\log s)[( \exp(- t)/s)^{2\Gamma} + {\rm Re}( b (\exp(-t)/s)^{2\Gamma+2i\Omega})]$ (solid).
   }
   \label{fig:gkz}
\end{figure} 

The SYK model has a unitary parity symmetry, $[H,P]=0$, where $P$ anticommutes with all $\gamma_i$. Because of this symmetry, the Hamiltonian can be block-diagonalized into two parity blocks, whereas the Majorana operators $\gamma_i$ are block off-diagonal. 
Additionally, the SYK Hamiltonian has an antiunitary symmetry $CK$, $[H,CK]=0$, where $C$ is unitary and $K$ is the complex conjugation operator \cite{garcia2016,cotler2016}.
We focus on the most interesting case $N \mod 8 = 2$ (class A).
In this case, $CK$ does not act as a symmetry in each parity block of the Hamiltonian \cite{garcia2016,cotler2016}; instead, the eigenvalues of both parity sectors are the same, with their eigenvectors related by $|\wtk\rangle \equiv C|k\rangle^*$. Similar results are obtained for other values of $N \mod 8$, which we defer to a separate publication.

In Fig.~\ref{fig:all} we show the Green's function, $G(t)$, and the OTOC,
$F(t)$, for $N=18$ and compare them to the analytical expressions derived below
(see Appendix \ref{app:N26} for $N=26$).
We can distinguish five different domains that we will discuss in detail below.
1.\ The perturbative domain, $0 < t < 4$ (purple).  
2.\ The scrambling domain, $1< t < \log N$ (orange, only for the OTOC). 
3.\ The relaxation domain, $ \log N < t < N$ (green). 
4.\ The power-law domain, $N < t <  D^{1/3} \exp(-\alpha N/3)$ (blue), where $\alpha=\pi^2/(4q^2)$.
5.\ The ergodic (RMT) domain,  $ t  > D^{1/3} \exp(-\alpha N/3)$ (red).

A plausible simplified RMT description of the late-time dynamics is through a Hamiltonian for which the SYK eigenvectors are replaced by Haar-random eigenvectors, while keeping the SYK spectrum and symmetries; we refer to this model as SYKGUE.
As is clear from Fig.~\ref{fig:all} (gray curves), the SYKGUE
does not, however, quantitatively describe the SYK data at all time scales, starting from the perturbative domain.

{\it Perturbative domain.---}%
The perturbative expansion of the Green's function and OTOC is an expansion in powers of $t^2$ for the SYK, while it is an expansion in powers of $t^2N$ for the SYKGUE, which departs
at a much faster rate from its initial value (compare the gray and purple curves in Fig.~\ref{fig:all}).
This can already be seen from the  coefficient of $t^2$ of $G(t)$, which can be computed as
\begin{equation}
2\sigma_E^2 =\frac 1D \sum_{kl} \overline{(E_k-E_l)^2 |\langle k|\gamma_0|l\rangle|^2} = -\frac 1D \tr \overline{[H,\gamma_0]^2 },\!\!\!
\end{equation}
where the overline denotes the ensemble average and the sum runs over both parity sectors.
For the SYK model the commutator gives
$
\sigma_E^2
\sim  N^0,
$
while for the SYKGUE the average over the eigenvalues and eigenvectors factorizes, resulting in
$
\sigma_E^2
\sim N/q.$
The width of the distribution for the SYK is a factor $\sqrt{q/N}$ smaller than for the SYKGUE, resulting in a peak value that is enhanced by a factor $\sqrt{N/q}$. We will see that the same mechanism is responsible for the enhancement of the SYK compared to the SYKGUE in the RMT domain (Fig.~\ref{fig:all}).
The perturbative expansion of $G(t)$ can be resummed
into a sum of decaying isolated RP resonances \cite{Dodelson:2024atp}; while expected, a similar calculation has not yet been carried out for the OTOC.

{\it Scrambling and RP resonances.---}%
Explicit results for the OTOC beyond the scrambling time are known in the large-$q$ limit,
by using the scramblon formalism \cite{gu2022}, which is applicable up to time scales $\exp \lambda_L t \sim N$, where $\lambda_L$ is the Lyapunov exponent.
For large $q$, the resulting asymptotic
series in powers of $z =(\exp \lambda_L t)/N$ could be resummed, with leading decay rates  equal to twice those of the retarded Green's function \cite{maldacena2016}.  
Since it is not clear whether this behavior \cite{polchinski2015} persists for finite $q$,
we will address this question for $q=4$.

In the scramblon formalism, the OTOC is obtained by solving the SD equations with an out-of-time-order source, which yields the Wightman function $G_W(t)$  \cite{gu2022}. 
The Wightman function can be expressed as the Laplace transform of a function $h(y)$ \cite{gu2022}, $G_W(u) =\int_0^\infty dy e^{-uy} h(y)$, from which the OTOC follows:
\begin{equation}
\label{otoc}
F(u=s e^{\lambda_L t})   = -\int_0^\infty\!\!\!\int_0^\infty dy_1 dy_2 e^{-u y_1 y_2}  h(y_1)h(y_2).
\end{equation}
Here, $u = s e^{\lambda_L t}$ is the scaling variable, and the source strength $s$ is proportional to $1/N$. The proportionality constant is fixed by comparing to the first term of the scramblon expansion \cite{gu2019a}, which gives $ s = 0.90/N$. 

In general, $G_W$ can only be computed by numerically solving the SD equations. Calculating the OTOC with Eq.~(\ref{otoc}) then requires a generally ill-conditioned Laplace inversion of the numerical $G_W$. 
To avoid this step, we fit the numerical $G_W$ by an ansatz based on the
shockwave expansion for the Schwarzian limit~\cite{maldacena2016a,stanford2022,Liu:2026nnw} with the necessary modification due to the complex nature of the RP resonances. This allows us to evaluate the inverse Laplace transform analytically and therefore compute the OTOC.
We require this ansatz to reproduce the initial exponential growth of the OTOC in the small-$u$ limit and the exponential decay at long times (large $u$). Such an ansatz is given by (see Appendix \ref{app:an})
\begin{equation}
   G_W(u)= \frac{a_1}{(1/2+  u)^{2\Gamma/\lambda_L}}
   +       2 {\rm Re}\left [\frac{a_2}{(1/2+  u)^{2(\Gamma+i\Omega)/\lambda_L}}\right ].
   \label{gwan}
\end{equation}
Here, $\Gamma \pm i\Omega$ are the leading RP resonances of the retarded Green's function~\cite{garcia2022e,Dodelson:2024atp} and $a_1\in\mathbb{R}$, $a_2\in\mathbb{C}$. 
Two of the three parameters $a_1$, $a_2$, and $a_2^*$ are fixed by the values of $G_W(0)=1/2$ and $G_W'(0)$ (obtained from the initial growth of the numerical $G_W$), leaving only one free (fitting) parameter.
In Fig.~\ref{fig:gkz} (top), we compare the solution of the sourced SD equations (black dots) to this ansatz, which gives an excellent fit to the data.

In Fig.~\ref{fig:gkz} (bottom) we show the result for the OTOC obtained from the scramblon formalism based on the ansatz Eq.~\eref{gwan}.
Its tail (red) is the sum of purely decaying terms $\sim  (\lambda_L t+\log s) \exp(- 2\Gamma t)$  (dashed red) and decaying
oscillatory terms $\sim( \lambda_Lt +\log s)\exp (-2(\Gamma  \pm i \Omega) t)$, but the
oscillations are weaker than  those of $G_W(t).$ Also shown is $G^2(t)$ (green curve),
which is proportional to ${\rm Re} \exp (-2(\Gamma  \pm i \Omega) t)$. It is much smaller
than $- F(t)$ because the latter behaves as  $1/4  - s \exp \lambda_L t$ before it starts decaying. 
In Fig.~\ref{fig:all} (bottom) we compare the OTOC computed with Eq.~\eref{otoc} (orange) to the numerical $N=18$ result (black); good agreement is found until well into the relaxation domain.
The growth domain ends when $\exp(\lambda_L t)/N \approx 1$, beyond which the scramblon expansion becomes questionable.
Nevertheless, if the OTOC for $t \gtrsim \log N$ is a sum of resonances, they will be  determined by times  $t \lesssim \log N$, which explains its extended validity.
Note also that the asymptotic expansion in
powers of $z$ is a double-scaling limit. 
Indeed, at least in the low-temperature limit, the scramblon formalism agrees with the
gravitational calculation \cite{maldacena2016a} even in the $z \gg 1$ limit provided that $t \propto \log N$. 
   
\begin{figure}[t]
	\includegraphics[width=0.74\columnwidth]{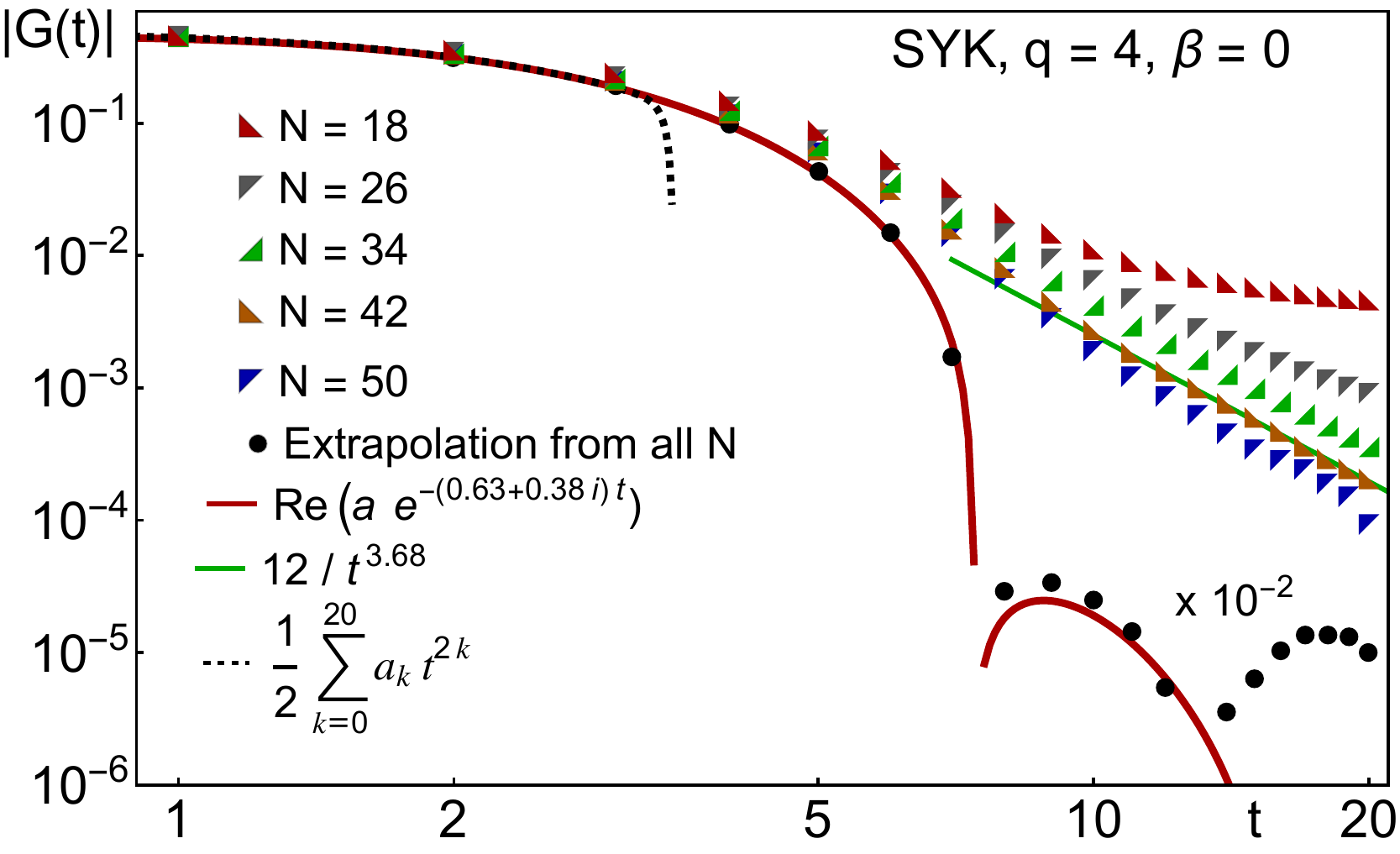}
	\caption{Finite-$N$  Green's function Eq.~(\ref{eq:green}) versus time for various $N$, obtained by exact diagonalization and Krylov subspace techniques for $\beta = 0$. The black dots are the quadratic extrapolation in $1/N$ from all values of $N$. They are compared to the leading resonance result of Ref.~\cite{Dodelson:2024atp} including the prefactor (red curve) and the perturbative result (black dashed curve). The extrapolation for $t > 7.5$ is multiplied by $0.01$ to avoid
	overlap with the finite-$N$ data. For $N=42$, we show a fit of the power-law domain (green line).  
	}
	\label{fig:gpowerlaw}
\end{figure}   
   
{\it Power-law domain and finite $N$.---}%
The results discussed up to now have a well-defined large-$N$ limit. To understand the approach to the
thermodynamic limit, we have studied the $N$-dependence of the Green's function, see Fig.~\ref{fig:gpowerlaw} for $18\le N\le 50 $ with $N\mod 8 =2$ at times that reach well into the power-law domain ($t >10$).
We also show the results of a quadratic extrapolation in $1/N$ (black dots). The power-law tail does not
survive the extrapolation, confirming that it vanishes for $N\to\infty$; indeed, it is due
to the square root tail of the spectral density, which is suppressed as
$\exp(-\alpha N) $ and gives a $1/t^3$ tail for the Green's function, and a  $1/t^6$ tail for the OTOC.
The extrapolated result is compared
to the result obtained from solving the SD equations (red curve) including the overall prefactor. It also agrees
with the perturbative expansion of the Green's function \cite{Dodelson:2024atp}
up to the convergence radius of $t \approx 4$ (black dashed curve).

\begin{figure}[t]
	\includegraphics[width=0.78\columnwidth]{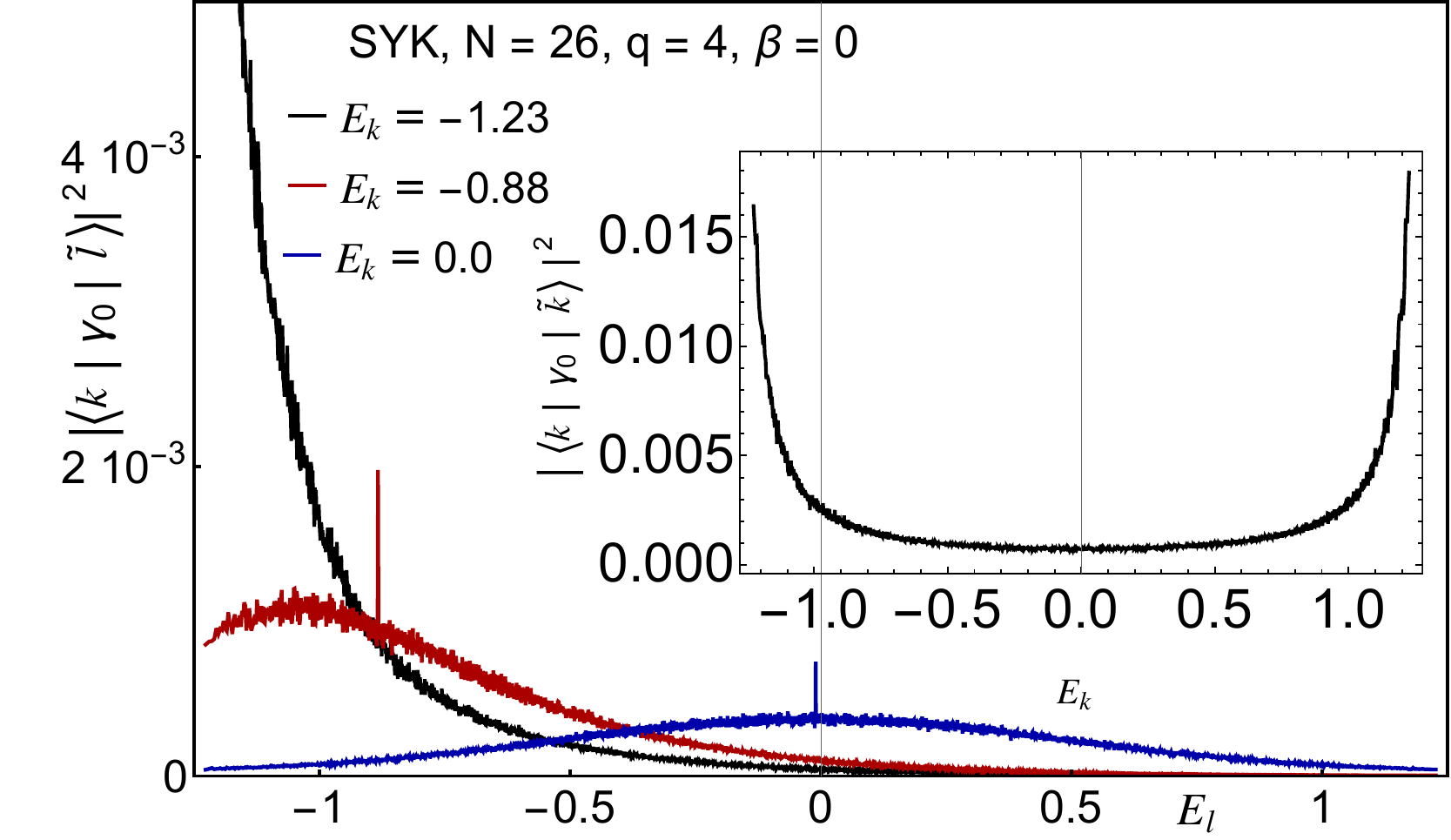}
	\caption{Energy dependence of the average matrix elements $|\langle\wtk |\gamma_0|l\rangle |^2$ as a function of $E_l$ for fixed $E_k$. Note the enhancement by a factor 2 for $k=l$. The diagonal matrix elements $\langle|\wtk |\gamma_0|k\rangle|^2$ (inset) only show a weak dependence on the energy in the bulk of the spectrum. The matrix elements for $k \approx l$, which give the main contribution to the Green's function, are much larger than the SYKGUE values of $(1+\delta_{kl})/D$.
        }
	\label{fig:matrixele}
\end{figure}

The exponent of the power-law increases with $N$, but our data
are not sufficient to extract its large-$N$ limit. Contrary to the power-law decay in RMT \cite{torresherrera2017,cipolloni2024}, the SYK result does not show oscillations, and is enhanced by a large $N$-dependent factor, which has also been found in other many-body quantum chaotic systems \cite{torresherrera2017}
as well as both in the Schwarzian SYK \cite{mertens2017,bagrets2017,mertens2023} and the DSSYK \cite{berkooz2019,berkooz2024}. In both these cases, the matrix elements of the Majorana operator depend on energy, resulting in a vastly enhanced coefficient of the $1/t^3$ tail.
This is also the case for the SYK \cite{Sonner:2017hxc} (Fig.~\ref{fig:matrixele}), while in RMT the average matrix elements are energy independent because of the independence of
eigenvalues and eigenvectors. The power-law domain starts when it dominates the RP resonances, i.e., at $ t \sim \alpha N/\Gamma$, and it ends when the dip value becomes dominant, i.e., at $ t \sim D^{1/3} \exp(-\alpha N/3)$. 
 
{\it  Late stages of many-body quantum chaos and RMT.---}%    
The discreteness of the spectrum, which is washed out in the large-$N$ limit, becomes important for time scales as early as $t\sim \log N$.
Although the SFFs of the SYK and SYKGUE coincide (since the SFF is fully determined by the eigenvalues) \cite{garcia2016,cotler2016}, we have already seen that, quantitatively, this is not the case for the Green's function and the OTOC (Fig.~\ref{fig:all}, gray curves). In the SYKGUE, we replace the SYK eigenvectors by Haar-random eigenvectors; averaging over the eigenvectors
\cite{Creutz:1978ub,weingarten1978asymptotic,collins2006,
garcia2016,cotler2016,cotler2020spectral,collins2022,Miesch:2023hjt} 
(Appendix~\ref{app:haar}) results in
 \be
\label{eq:gf}
\overline{G(t)} &=&{(K(t)/2 +1)}/ {(D+2)},\\
\overline{ F(t)}
    &=&
    b_1 K(t) + b_2 K(2t) + b_3 {\rm Re} K_3(t) + b_4 K_4(t) + b_5,\nn
    \ee
    where $b_k$ are given in Appendix~\ref{app:haar} and depend only on $D$. Here and below, we have explicitly denoted the disorder average (i.e., over energies and eigenvectors) by an overline. The SFF of the SYK spectrum is defined as
\be
K(t) =    \frac 4{D} \sum_{k  l} \overline{e^{it(E_k-E_l)}},
\label{sff}
\ee
and $K_{3,4}$  are  the 3- and 4-point SFFs (see Appendix \ref {app:haar}).
These results do not apply to the SYK because 
the matrix elements $\langle \wk |\gamma_0|l\rangle$ depend on $E_l$ at fixed $k$. Notwithstanding, in the RMT domain, the SYKGUE and the SYK results only differ by a constant-in-time, but $N$-dependent, shift, which we explain next.

The Green's function can be decomposed as 
\begin{equation}
	\label{eq:G_RMT}
 \overline{G(t)} = \frac 2{D} \Big(\!\sum_{k\ne l} \overline{e^{it(E_k-E_l)} |\langle \wtk  |\gamma_0| l\rangle|^2}
   +  \sum_k   \overline{|\langle \wtk |\gamma_0| k \rangle |^2 } \Big)\!,\!\!
\end{equation}
where the second term gives the saturation value beyond the Heisenberg time (sums are over one parity sector). It is essential that the eigenvalues in both parity sectors are the same and that $\langle \wtk |\gamma_0| k\rangle $ is nonvanishing \cite{cotler2017},  which is only the case for $ N \mod 8 =2$ (see also Ref.~\cite{foini2025} for an example of symmetry-class dependence of the ramp-plateau structure in correlation functions). To derive an analytical result we make two assumptions: 1.\ For $t> t_{\rm dip}$, the main contribution to the Green's function is from pairs of eigenvalues that are close (\textit{locality in energy}). Otherwise the contribution is suppressed by a large phase. 2.\ For this limited range of energies, the matrix elements $\langle \wtk |\gamma_0| l \rangle$ do not depend significantly on $E_l$ for fixed $k$. However, this does not mean that $\langle \wtk|\gamma_0| l \rangle \approx \langle k |\gamma_0| k \rangle$ (because eigenvectors of opposite parity are related by $|\wtk\rangle  \equiv  C|k\rangle^*$).
In terms of wave functions $\phi_k$, we can write the matrix elements in Eq.~(\ref{eq:G_RMT}) as
\be
 |\langle \wtk |\gamma_0| k\rangle|^2  = (\phi_k^* \gamma_0 C \phi_k^* )( \phi_k \gamma_0 C \phi_k).
\ee
Since the components of the eigenvectors are Gaussian distributed, we have two equivalent contractions, while there is only one for $ k \ne l $.
This results in $\overline{ |\langle \wtk |\gamma_0 | l \rangle|^2_{k\ne l, k \approx l}}
 = \frac 12\overline{|\langle \wtk |\gamma_0| k \rangle |^2}$ (Fig.~\ref{fig:matrixele}).
The Green's function beyond the dip time thus reads
 \be
\overline{ G(t)}
&=& \frac 1{D}  \sum_{k  l} \overline{ (e^{it(E_k-E_l)} 
   +\delta_{kl} )    |\langle\wtk |\gamma_0| k \rangle |^2}.
\label{grrmt}
\ee
Both the SFF and the diagonal matrix elements depend on energy, but in the bulk of the spectrum, this dependence is weak. If we further make the stronger assumption that the averages over the eigenvectors and eigenvalues factorize, we obtain
   \be
   \overline{G(t)}
 &=& \frac 1{4} \left( K(t)
   + 2  \right )  \overline{ |\langle \wtk  |\gamma_0| k \rangle |^2} .
\label{gfinal}
   \ee
   The average value of the diagonal matrix element  $\overline { |\langle \wtk|\gamma_0| k \rangle |^2}$ can easily be calculated numerically, which gives us a parameter-free prediction for the enhancement of the Green's function in the RMT domain. This is the result shown by the red curve in Fig.~\ref{fig:all} (top). The result Eq.~\eref{gfinal} also explains why the value of the Green's function at the dip is half the saturation value beyond the Heisenberg time. The SYKGUE result \eref{eq:gf} is recovered by using
 Haar eigenvector moments \cite{mcdonaldaverages1972}, which gives $\overline { |\langle \wtk|\gamma_0| k \rangle |^2}= 2/(D+2)$.

A similar argument can be made for the OTOC, resulting in (Fig.~\ref{fig:all}, bottom)
     \be
     \overline{F(t>t_{\rm dip}) }\sim   \overline{G(t)}^2,
     \ee
     with proportionality constant determined by the matrix element $\overline{\langle k|\gamma_0|\wk\rangle\langle\wk|\gamma_1|m\rangle\langle m|\gamma_0|\wm\rangle\langle \wm|\gamma_1|k\rangle} $ (see Appendix \ref{app:RMTotoc}).  For the SYKGUE, to leading order in $1/D$, we obtain
     $  \overline{F(t>t_{\rm dip}) }=-2 \overline{G}^2(t)$ (see Appendix~\ref{app:RMTotoc}), which also
     follows from its exact finite-$D$ expression Eq.~\eref{eq:gf}.
The disagreement with the plain-RMT prediction after the dip time points to a lack of decoupling between eigenvalues and {\it matrix elements} of Majorana operators in SYK.
In the context of quantum gravity, deviations from RMT cast doubts on the range of applicability of the proposal \cite{saad2019,Altland:2024ubs} that, for sufficiently long times after the dip, the field theory dual to JT gravity with no matter is a random matrix. Further research is required to settle this issue.

{\it Conclusions.---}%
Employing the $q = 4$ SYK model, we have studied the main time scales of the dynamics of many-body quantum chaotic systems, as captured by the OTOC and the retarded Green's function. 
All analytical expressions agree well with numerical results obtained from exact diagonalization and Krylov subspace techniques.
The five domains of many-body quantum chaos are not independent. As shown in Ref.~\cite{Dodelson:2024atp}, the perturbative expansion reproduces the resonances. The $\sqrt{ N/q} $ suppression
of the leading-order coefficient of the perturbative expansion also leads to  the enhancement factor beyond the Heisenberg time. Finally, the leading decay rates of the OTOC are determined by  the leading decay rates of the Green's function. 
This interdependence shows that the OTOC, as a signature of many-body chaos, should be considered at all time scales.

An important problem that deserves further attention is the precise delimitation of the universality of these results. 
The scramblon approach to describe the OTOC is applicable to random spin models~\cite{liu2024zhang} and to long-range interacting spins in solid-state NMR~\cite{li2026zhang}. Exponential growth followed by RP relaxation is also found in systems with a chaotic classical limit~\cite{garciamata2018}. RP resonances and the dip-ramp-plateau should also be a generic feature of many-body quantum chaotic systems~\cite{das2025santos,yoshimura2025chalker,yoshimura2025theory} because of the need to relax to an ergodic state.
However, it would still be desirable to confirm this for spin chains or other fermionic systems beyond SYK. Lacking the analytic tractability of the SYK model, this will be a challenging numerical task.

{\it Acknowledgments.---}%
We thank Zhenbin Yang for useful discussions during the Simons Center Workshop  ``Random Geometry in Math and Physics''.
A.\ M.\ G.\ G.\ acknowledges support
from the Natural Science Foundation of China (NSFC)
through the Individual Grant
No.\ 12374138. L.\ S.\ was supported by a Research Fellowship from the Royal Commission for the Exhibition
of 1851. J.\ J.\ M.\ V.\ is supported in part by U.S. DOE
Grant No.\ DE-FG02-88ER40388, and acknowledges hospitality and support from Jiaotong University.

\bibliography{libotoclind3.bib}
\appendix

\section*{END MATTER}

\section{Green's function and OTOC for $N=26$}
\label{app:N26}

\begin{figure}[t]
	\begin{tabular}{@{}r@{}}
	\includegraphics[width=0.75\columnwidth]{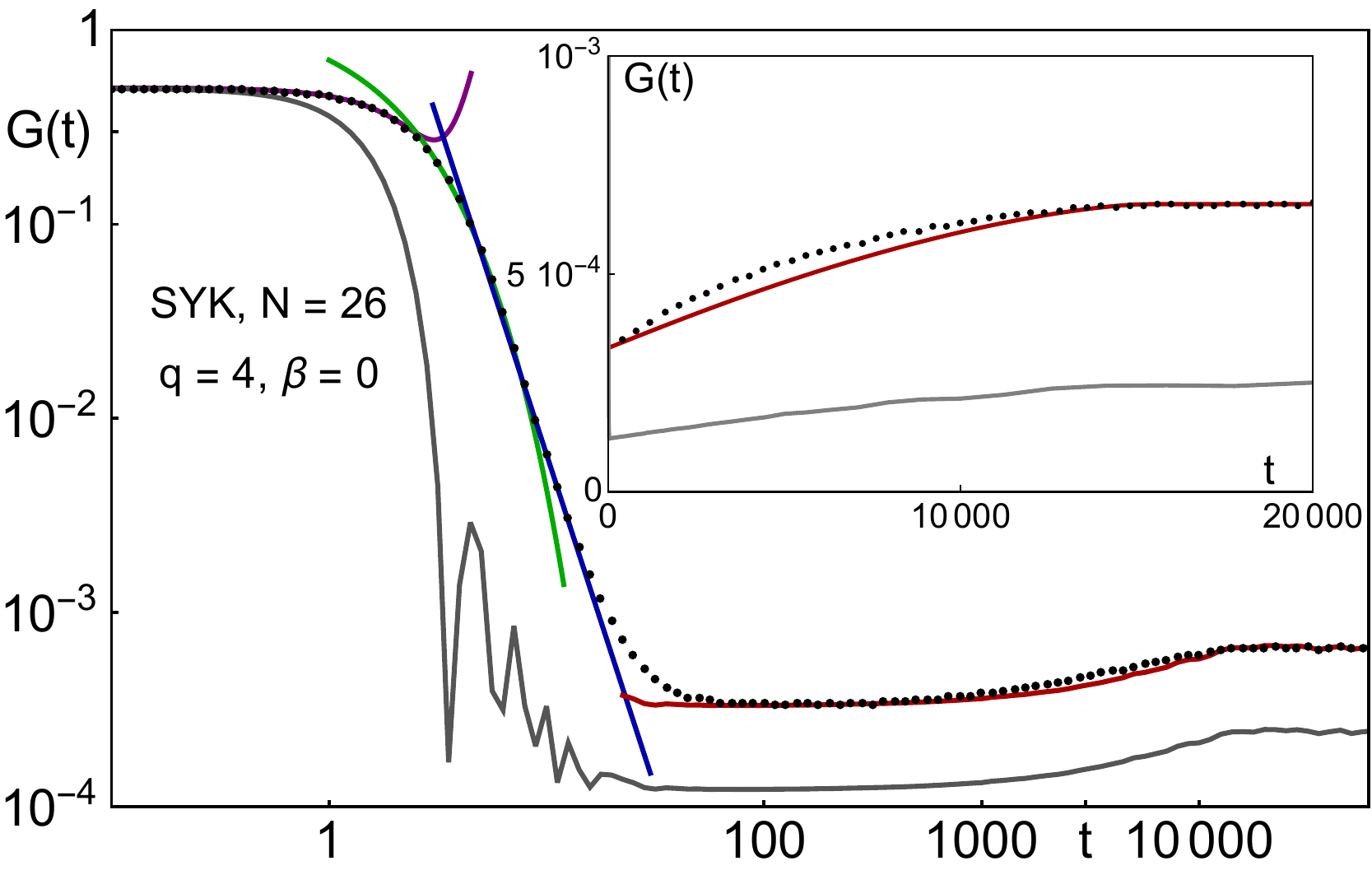}\\
	\includegraphics[width=0.76\columnwidth]{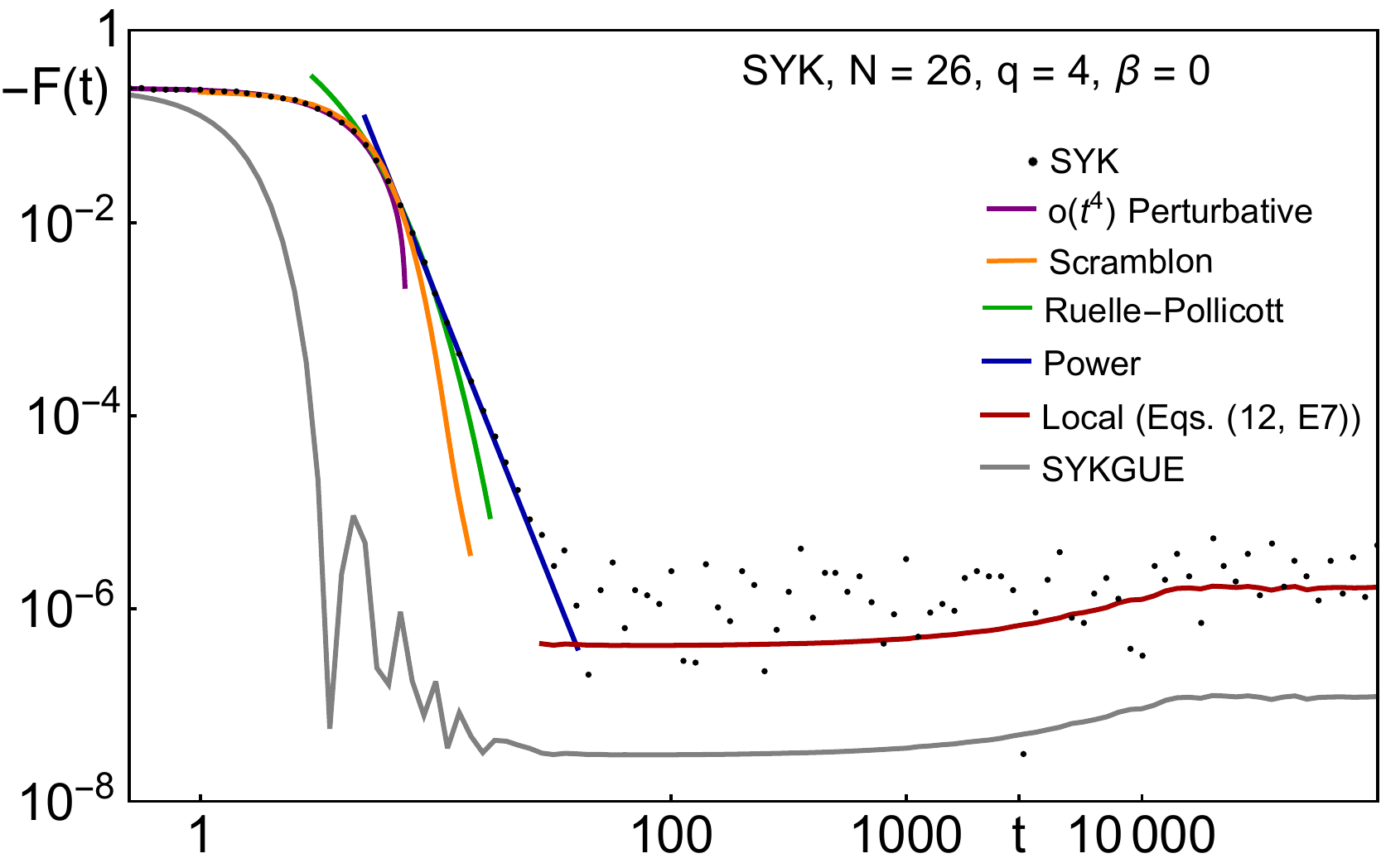}
	\end{tabular}
	\caption{Time dependence of the Green's function (top) and  OTOC (bottom) on a log-log scale. Numerical results (black dots) obtained from exact diagonalization for the $N=26,\; q=4, \; \beta=0$  SYK model are compared to the analytical results and numerical fits derived in the main text, given by solid curves as indicated in the legend of the bottom panel.
    The inset in the top panel figure shows the same curves for $t >100$ on a linear scale. The correction to the SYKGUE result is large, but because of the factorization assumption there is still a 10\% discrepancy between analytical and numerical results.
        }
        	\label{fig:all26}
\end{figure}

In Fig.~\ref{fig:all26}, we show the Green's function and OTOC for $N=26$. The fluctuations of the OTOC in the RMT domain are large (resolving its exponentially small signal requires a growing number of computationally expensive disorder realizations), but the overall picture is the same as in the $N=18$ case discussed in the main text.

\section{Numerical parameters}
\label{app:fit}

In this appendix, we list the numerical values of the parameters of the curves plotted in the figures.
Note that at finite $N$, there can be significant $1/N$ corrections. Another source of inaccuracy is that
at finite $N$ the oscillations of  the Green's function and the OTOC are suppressed, and it is not
possible to determine reliably the strength of this component. However, they affect the parameters
of the exponential and power-law decay.

For the perturbative curves in Figs.~\ref{fig:all} and \ref{fig:all26}  we use the exact finite-$N$ values. The SYKGUE result is the exact
finite-$N$ result with numerically determined SFFs. The local approximation is evaluated exactly,
with a numerically determined coefficient, which is the combination of matrix elements that gives the
enhancement factor, $\epsilon_h$,  beyond the SYKGUE result. This and the parameters of
$a/t^p$ and $b \exp( -\Gamma t)$ are shown in Table~\ref{app:tab}.
The parameters of the ansatz for $s=10^{-6}$ are equal to
$a_1=0.37$ and $a_2 = -0.0245+0.13 i$.

\begin{table}[t]
	\caption{Fitting parameters of $a/t^p$ and $b \exp(-\Gamma t)$ for the Green's function and OTOC and numerically calculated enhancement factor $\epsilon_h$ [see Eq.~\eref{D7}].}
	\begin{tabular}{cc|ccccc}
		& $N$ & $a$ &$ p$ & $b$& $\Gamma$ & $\epsilon_h$ \\
		\hline
		$G(t)$ & 18 & 16.5 & 3.2&1.2& 0.52 &1.96\\
		$G(t)$ & 26& 18&3.44&1.23&0.57&2.7\\
		$F(t)$ & 18& 480&5.45&2.2 & 0.75&1.92\\ 
		$F(t)$ & 26& 2300&6.1&3.1& 0.75&13.6\\
	\end{tabular}
	\label{app:tab}
\end{table}
        
\section{Ansatz for the Wightman function}
\label{app:an}

In Eq.~\eref{gwan}, we introduced the ansatz for the Wightman function, which in its general form is given by
\be
G_W(u=se^{\lambda_L t}) = \sum_{k}  a_k \frac 1{\left (1/2 +u/{u_k} \right)^{\alpha_k}}. 
\ee
For $\beta =0$, $u_k =1$, but this is not the case at nonzero $\beta$.
The exponents $\alpha_k$ and amplitudes $a_k$ can be complex, but $G_W(u)$ is real. The Laplace
decomposition of $G_W(u)$ is given by $G_W(u) =\int_0^\infty dy e^{-u y} h(y)$. For each of the terms we obtain
\be
h_k(y) = \frac{a_k e^{-u_k y/2} (u_k y)^{\alpha_k }}{y \Gamma(\alpha_k)},
\ee
where $\Gamma(z)$ is the gamma function.
Note that the small-$y$ limit of $h_k(y)$ is determined by the leading asymptotic 
behavior of $G_W(u)$. Following Ref.~\cite{gu2022}, the OTOC is
\begin{equation}
F(u=s e^{\lambda_L t} ) = -\int_0^\infty\!\!\!\int_0^\infty dy_1dy_2 e^{-u y_1y_2} h(y_1) h(y_2).
\end{equation}
The integrals can be evaluated analytically:
\begin{equation}
F(u) =- \sum_{kl} a_ka_l 
\frac{  U(\alpha_k,1\!+\!\alpha_k\!-\!\alpha_l,  u_ku_l/ {4u})}{ 2^{\alpha_l-\alpha_k}(u/u_k u_l)^{\alpha_k}},
\end{equation}
with $U(a,b,z)$ the Tricomi confluent hypergeometric (hypergeometric $U$) function.
Using the known asymptotics of the hypergeometric functions, for small $u$ we find
\begin{equation}
  F(u) = -\sum_{kl} a_k a_l 2^{\alpha_k+\alpha_l}  (1-4u/u_ku_l)+\cdots,
\end{equation}
while for large $u$ it is given by
\begin{equation}
  F(u) \sim -\sum_{kl} a_k a_l
  \left( \frac{ 2^{\alpha_l-\alpha_k}\Gamma(\alpha_k-\alpha_l)}{  (u/u_ku_l)^{\alpha_l}\Gamma(\alpha_k)} + (k \leftrightarrow l )  
  \right)
\end{equation}
for $k \ne l$, and
\begin{equation}  
        F(u) \sim -\sum_k a_k^2 \frac {\log u - \Gamma'(\alpha_k)/\Gamma(\alpha_k)- 2 \gamma_E}{\Gamma(\alpha_k) u^{\alpha_k}}
\end{equation}
for $k=l$, with $\gamma_E $ the Euler constant. The term $\sim \log u$ follows from the leading asymptotics of $h(y)$ [or $G_W(t)$].

\section{Calculation of Green's function and OTOC by Haar integration}
\label{app:haar}

We now consider the SYKGUE for $N \mod 8 =2$, where the eigenvectors
are distributed according to the Haar measure, with the eigenvectors of the two parity sectors related by charge conjugation. Performing the Haar integrals
\cite{Creutz:1978ub,weingarten1978asymptotic,collins2006,collins2022,garcia2016,cotler2016,cotler2020spectral,Miesch:2023hjt}
we
obtain the  OTOC in Eq.~\eref{eq:gf}. The constants $b_k$ are given by
(for $N  \mod 8 =2$)
\be
\label{bk}
b_1&=&
-8 ( D^5-6D^4-20 D^3+120 D^2+64  D-384)/D_W,
\nn    \\
b_2&=&
-4 (D^4-8 {D}^3+12 {D}^2-32 {D}+192)/D_W,
\nn \\
b_3&=&
32 ({D}^3-8{D}^2+4 {D}+48 )/D_W,
\nn \\
b_4&=&
-D({D}^4-8 {D}^3+12 {D}^2-32 {D}+192)/D_W,
\nn    \\
b_5&=&-8D^2 ({D}^3-4  {D}^2-36 {D}+144 )/D_W,
\ee
with Weingarten denominator $D_W=4D (D^2-36)(D^2-16)(D^2-4)$.
The 3- and 4-point correlation functions in Eq.~\eref{eq:gf} are defined as
\be
\label{eq:def_K34}
K_3(t)
&=&\frac{8}{D}\sum_{i,j,k=1}^{D/2}\overline{e^{-i(2E_i-E_j-E_k) t}},
\nn\\
K_4(t)
&=&\frac{16}{D^2}\sum_{i,j,k,l=1}^{D/2}\overline{e^{-i(E_i+E_j-E_k-E_l) t}}.
\ee
They are normalized as $K_3(0)=K_4(0)=D^2$, and $\lim_{t\to\infty}K_3(t)=4$, $\lim_{t\to\infty}K_4(t)=8$. 
For $t\gg 1$, we can replace $K_4(t)\approx 2K(t)^2$ and recover result
$\overline{F(t)}=  -2\overline{G(t)}^2$ to leading order, as quoted in the main text.

\section{OTOC in the RMT domain}\label{app:RMTotoc}

The OTOC can be expanded as
\begin{equation}
 \overline{F(t)} =\frac 2D\sum_{klmn} \overline{e^{it(E_k-E_l) }e^{it(E_m-E_n)} f(k,l,m,n)},
\end{equation}
where
\begin{equation}
f(k,l,m,n) = \langle k | \gamma_0|\wl\rangle \langle \wl | \gamma_1| m\rangle
 \langle m |\gamma_0| \wn\rangle \langle \wn | \gamma_1|k\rangle,
 \end{equation}
and the sum is over one parity sector; the other sector gives the same result. We can make an argument for the OTOC similar to the one for the Green's function in the main text. Assuming locality in energy and additional contractions for diagonal matrix elements, we obtain
   \be
   \overline{f(k,l,m,n)} = \frac 14 \overline{f(k,k,m,m)},\nn\\
   \overline{f(k,k,m,n)} = \frac 12 \overline{f(k,k,m,m)}.
   \ee
Making the same assumptions as for the Green's function and splitting the sums into sums over equal and different energies, we find
      \be
      \overline{F(t)}
&=&\frac 1D
\sum_{km} \overline{f(k,k,m,m)} \sum_{ln}
(\overline{e^{it(E_k-E_l) }} +\delta_{kl} )
\nn\\ &&\times
( \overline{e^{it(E_m-E_n)}} +\delta_{mn}).
\ee
For the SYKGUE, $ \overline{f(k,k,m,m)} = -8/D^3$ to leading order. This results in the
OTOC
\be
\overline{F_{\mathrm{SYKGUE}} (t)} = -\frac 1{2D^2} (K(t) + 2)^2.
\label{C5}
\ee
If we assume that the energy dependence of $\overline{ f(k,k,m,m)}$
is weak, the OTOC for the SYK can be approximated as
\be
\overline{F_{\mathrm{SYK}}(t)}
&=& -\frac {D^3}8   \overline{f(k,k,m,m)}\, \overline{ F_{\mathrm{SYKGUE}}}(t)
\equiv \epsilon_h \overline{ F_{\mathrm{SYKGUE}}}(t).\nn\\
\label{D7}
\ee
(see Table \ref{app:tab} for numerical values of the enhancement factor $\epsilon_h$).
Using the result for the Green's function, we thus have
\be
\overline{F_{\mathrm{SYK}}(t)} = D \frac{\overline {f(k,k,m,m)} }{\overline{ |\langle \wtk| \gamma_0| k \rangle|^2}^2} \overline{G_{\mathrm{SYK}} (t)}^2.
\ee

\clearpage

\end{document}